\documentclass[twocolumn,prb, showpacs, amsmath, amssymb, amsthm]{revtex4}
\usepackage{bbm}
\usepackage[hang,loose]{subfigure}
\usepackage[dvipdfm]{graphicx}
\usepackage[ps2pdf,bookmarks,urlcolor=red,citecolor=blue,linkcolor=red,
               pagecolor=red,hyperindex,colorlinks,hyperfigures 
          ]{hyperref}

\renewcommand{\vec}[1]{\mathbf{#1}}

\begin{document}

\title{Exterior optical cloaking and illusions by using active sources:
a boundary element perspective}

\author{H. H. Zheng,$^{1}$ J. J. Xiao,$^{1,2,3}$ Y. Lai,$^{1}$ and C. T.
Chan$^{1}$}
\affiliation{$^1$Department of Physics and William Mong Institute of
NanoScience and Technology, The Hong Kong University of Science and
Technology, Clear Water Bay, Hong Kong, China\\
$^2$Department of Electronic and Information Engineering, Shenzhen Graduate
School, Harbin Institute of Technology,  Shenzhen 518055, China\\
$^3$Key Laboratory of Network Oriented Intelligent Computation, Shenzhen
Graduate School, Harbin Institute of Technology, Shenzhen 518055, China}

\begin{abstract}
Recently, it was demonstrated that active sources can be used to cloak any objects
that lie outside the cloaking devices [Phys.~Rev.~Lett.~\textbf{103},~073901~(2009)]. Here, we
propose that active sources can create illusion effects, so that an object outside the cloaking 
device can be made to look like another object. Invisibility is a special case in which the 
concealed object is transformed to a volume of air. From a boundary element perspective, we show
that active sources can create a nearly ``silent" domain which can conceal any objects inside and
at the same time make the whole system look like an illusion of our choice outside a virtual boundary.
The boundary element method gives the fields and field gradients (which can be related to monopoles 
and dipoles) on continuous curves which define the boundary of the active devices. Both the cloaking 
and illusion effects are confirmed by numerical simulations. 
\end{abstract}

\date{\today}
 \maketitle
\section{Introduction}
The classical wave scattering cross section of an object can be significantly
larger or smaller than the geometric cross section.~\cite{Kerker75,
Alu05} Recipes to achieve invisibility (zero cross section) are particularly
intriguing,~\cite{Dolin,Greenleaf2,Leonhardt06,LeonhardtNJP,
Pendry06,Schurig06, PendryLi,Pendry09} normally enabled by the concept of
transformation media and artificial
metamaterials.~\cite{Shalaev08,Pendry06,Enoch09,Zhang09} The correspondence between 
coordinate transformation and material parameters was noted nearly half a century ago ~\cite{Pendry06, Dolin}
and such correspondence was explicitly formulated as the technique of ``transform optics" to achieve 
invisibility by steering electromagnetic waves around a domain.~\cite{Pendry06, Schurig06} Similar
approaches to achieve invisibility were also proved mathematically for geometric optics ~\cite{Leonhardt06, LeonhardtNJP}
and in the quasi-static limit.~\cite{Greenleaf2} 
These invisibility devices typically work by steering light around an object and the material shells 
need to encircle the object to be cloaked.~\cite{Pendry06} It
was then proposed that ``cloaking at a distance'' can be achieved,~\cite{Lai09} and the
concept of cloaking can be extended to create arbitrary illusions.~\cite{Lai09_2} However, these
recipes are based on artificial metamaterials and usually have bandwidth limitations. 
Recently, cloaking by using active sources have been proposed,~\cite{Miller06, Milton06, Nicorovici07} which
removes the requirement of metamaterials as well as the bandwidth limitation. 
Miller gave a detailed prescription to perform active source cloaking, and considered the case
in which the sources encircle the cloaked domain. 
It was then shown by Vasquez, Milton, and D. Onofrei, ~\cite{MiltonPRL} that exterior cloaking can be also realized by using several points (disks) of active multipole sources placed around the object to be cloaked.
This cloaking effect has been demonstrated numerically in a broadband fashion.~\cite{MiltonPRL,MiltonOE}

In this paper, we will employ a boundary element formulation to show that arbitrary illusions can 
be achieve using simple active sources, i.e. sources of fields and field gradients (monopoles and dipoles), placed on continuous curves; and active source remote cloaking is a special case. When the active sources are tuned properly according to incoming waves, a nearly ``silent'' domain is created such that any objects can be hidden inside. At the same time, invisibility is achieved by reducing the ``scattered'' waves of the active sources to be almost zero on a virtual boundary which encloses the whole system. We also extend this scheme of external active cloaking to create arbitrary illusions, by tuning the active sources to produce the same ``scattered'' waves on the virtual boundary as those scattered by the object chosen for illusion under the same incident wave. The physics of the cloaking and illusion effects by active sources can be understood clearly from the boundary element perspective.~\cite{Xiao08}

\section{Boundary Integral Method}
\begin{figure}[t]
\centerline{\includegraphics[width=0.45\textwidth,clip]{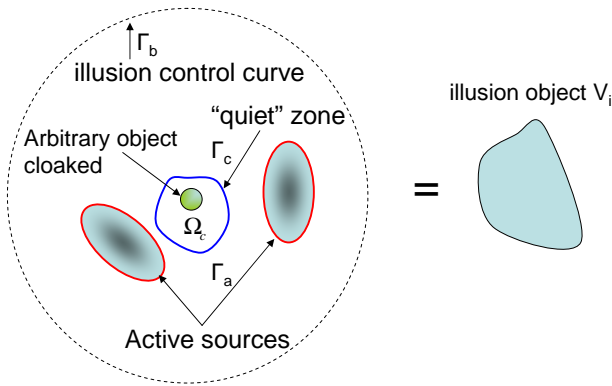}}
\caption{(Color online) Schematic figure for the illusion devices containing
exterior active sources on the boundaries $\Gamma_a$ (red lines) that
create a cloaked region $\Omega_c$ which can hide arbitrary scatterers inside and at the same time
make the total system response like another illusion scatterer $V_i$ outside a virtual
boundary $\Gamma_b$. When the illusion scatterer is just free space, invisibility is achieved.} \label{fig1}
\end{figure}

In this section, we demonstrate that the problem of two-dimensional (2D) active cloaking can
be formulated by using a boundary integral equation~\cite{Xiao08} and be extended to create arbitrary illusions.
A schematic figure of the configuration is shown in Fig.~\ref{fig1}.
The active sources are placed on boundary of the domains marked by the red lines (labeled as $\Gamma_a$). These active sources will generate fields so that any object inside a certain domain $\Omega_c$ will become invisible and the external observer (outside a virtual boundary $\Gamma_b$) will see an illusion of another object inside. The active sources do not need to
encircle the object to be cloaked, as shown in Fig.~\ref{fig1}. For any incoming waves, these active sources generate
opposite fields that cancel the incoming waves inside the domain $\Omega_c \subset R^2$ to make the total fields inside $\Omega_c $ almost zero. Any
objects inside this ``quiet" zone will be concealed as the objects experience no incident wave and thus no scattering will occur.

At the same time, the active sources can be used to generate outgoing fields outside the boundary
$\Gamma_b$ that mimic those scattered from another object $V_i$ under the illumination of the
same incoming wave, rendering the whole system to appear like the object
$V_i$ for any observers outside $\Gamma_b$. In other words, the wavefronts on
$\Gamma_b$ of the scattering fields from the object $V_i$ are reconstructed by
the active sources. Thus, the active sources can create an illusion so that any object placed inside
$\Gamma_c$ is transformed optically so that it looks like another object for observers outside $\Gamma_b$.
If the fields generated by active sources cancel each other outside a virtual boundary (labeled as $\Gamma_b$), then any observer
outside $\Gamma_b$ would not see the object inside $\Gamma_c$ as well as the active sources. In other words, the object inside $\Omega_c$ is cloaked by the active sources.
In this sense, invisibility is a special case in which
the illusion object $V_i$ is just free space.

The pre-designed object $V_i$, which is the illusion we want to
create, can be changed on demand as long as we can compute the
scattered fields of this object under the incoming probing waves.
In this active source approach, the external illusion device does not need to be tailor made for the object
and illusion, in contrast to the exterior cloak built with passive metamaterials,~\cite{Lai09_2} and
there is no intrinsic bandwidth limitation. However, one needs to know
in advance the probing wave, or one must set up sufficiently fast-responsive
sensors to capture the information of the probing wave on the
boundary of $\Omega_c$.~\cite{Miller06}

We will formulate the problem using the boundary element
approach. We note that the surface integral equation (SIE) of the 2D
Helmholtz equation $(\nabla ^2+k^2)\phi ({\rm {\bf r}})=0$, where $k$ is the
wave number, can be written as~\cite{Xiao08}
\begin{equation}
\label{eq:solinsidejin} \phi({\vec r})\Big|_{{\vec
r}\in\Omega}=\oint_{\partial \Omega}ds \big[g({\vec s}, {\vec r})
\partial_{\vec n}\phi({\vec s})-\phi({\vec s})\partial_{\vec n} g({\vec
s},{\vec r})\big]\,.
\end{equation}
Here ${\rm {\bf n}}$ is the outward unit normal vector on the boundary and
$\partial_{\rm {\bf n}}$ represents the normal gradient. The 2D SIE tells us
that inside a homogeneous domain $\Omega $, the scalar wave function $\phi
({\rm {\bf r}})$ is completely determined by the fields and the normal
derivatives on the boundary, connected by the Green's function $g({\rm {\bf
r}},{\rm {\bf {r}'}})=\frac{i}{4}H_{0}^{(1)}(k|\vec r-\vec r'|)$, where $H_{0}^{(1)}$ is the zeroth order of the first kind of Hankel function. The counterpart of Eq.~\eqref{eq:solinsidejin} for an
open domain which might also involves an incident field $\phi
^{\text{inc}}({\rm {\bf r}})$, reads $\phi ({\rm {\bf r}})\vert _{{\rm {\bf
r}}\in \left\{ {R^2-\Omega } \right\}} =\phi ^{\text{inc}}({\rm {\bf
r}})+\phi ^{\text{bc}}({\rm {\bf r}})$, where $\phi^{bc}$ is contributed
from the following boundary integral
\begin{equation}\label{eq:solinsidej}
    \phi^{\text{bc}}(\vec r)=-\oint_{\partial \Omega}ds
    \big[g({\vec s}, {\vec r}) \partial_{{\vec n}}\phi({\vec s})-\phi({\vec s})
    \partial_{\vec{n}} g({\vec s},{\vec r})\big]\,.
\end{equation}

Inspired by Eq.~\eqref{eq:solinsidej}, one can
construct additional fields ($\phi ^{\text{bc}})$ in the domain $R^2-\Omega$, by
appropriately choosing active sources to create $\phi ({\rm {\bf s}})$ and
$\partial _{\rm {\bf n}} \phi ({\rm {\bf s}})$ on the boundary $\partial \Omega$. In the Appendix,
we show that such kinds of active fields can be generated by a variety of sources either located on the boundary or inside the boundary and these fields correspond to
outgoing multipole radiation fields from a perspective of an observer outside the cloaking devices.
Using Eq.~\eqref{eq:solinsidej}, one can determine each component of the
multipole sources. Now, the issue is that whether it is possible to use such kind
of active sources to construct fields which can cancel the incident field
$\phi ^{\text{inc}}$ inside the ``quiet'' zone $\Omega _c $, and simultaneously
mimic the scattered field in the region outside $\Gamma_b $ (i.e., $R^2-\Omega _b )$, i.e.
\begin{eqnarray}\label{eq:condition1}
  \phi^{\text {bc}}(\vec r) = \left\{
  \begin{array}{r l}
    -\phi^{\text{inc}}(\vec r), &\quad \forall  \quad \vec r\in \Omega_c\,, \\
     \phi^{\text{sc}} (\vec r), &\quad \forall \quad \vec r \in {\mathbb R}^2-\Omega_b\,,
  \end{array}\right.
\end{eqnarray}
such that the total field is essentially zero inside $\Omega _c $, while outside
$\Omega_b $ the total field mimics the superimposition of the incoming waves
and the scattered wave of a pre-designed illusion object. As will be shown in Appendix ~\ref{Sec:Error}, the perfectness of the cloaking and illusion effect depends on the number of active sources that we can afford to use. Employing a uniqueness theorem~\cite{JAKong}, the conditions in
Eq.~\eqref{eq:condition1} can be simplified. In general, fixing the values of $\phi$ on the boundary
({\it{Dirichlet boundary condition}}) can already guarantee a unique
solution in the enclosed domain for the 2D Helmholtz equation. Thus, the
constraints in Eq.~\eqref{eq:condition1} can be replaced by
\begin{eqnarray}\label{eq:condition2}
  \phi^{{\text bc}}(\vec r) = \left\{
  \begin{array}{r l}
    -\phi^{\text{inc}}(\vec r), &\quad {\text {for}} \quad \vec r\in \Gamma_c\,, \\
    \phi^{\text {sc}} (\vec r), &\quad {\text {for}} \quad \vec r \in \Gamma_b\,.
  \end{array}\right.
\end{eqnarray}

In addition to the forgoing conditions, there is a self-consistent condition
of the boundary fields on $\Gamma_a$, ~\cite{Xiao08}
\begin{eqnarray}
  \frac{1}{2}\phi(\vec s) = \phi^{\text{inc}}(\vec s) &-& \int_{\partial
  \Omega_a}ds'\Big\{g(\vec s', \vec s)\partial_{\vec n}\phi(\vec s')\nonumber\\
  &-&\phi(\vec s')\partial_{\vec n}g(\vec s', \vec s)\Big\},\quad \vec s, \vec s'
  \in \Gamma_a\,.
\end{eqnarray}
in which, the integral is of a Cauchy principal value (CPV). This self-consistent
condition comes from the continuity requirement when  $\vec r$ approaches to the
boundary from outside the cloaking device, i.e.
\begin{eqnarray}
\lim_{\vec r\to \vec s}\phi(\vec r)=\phi(\vec s), \quad \vec r\in R^2-\Omega_a,
\vec s\in \Gamma_a\,.
\end{eqnarray}
In summary, the active fields are determined by the following integral equations,
\begin{widetext}
\begin{subequations}\label{eq:conditions_total}
\begin{eqnarray}\label{eq:condition_self}
  \frac{1}{2}\phi(\vec r) + \int_{\partial\Omega_a}\left\{g(\vec s, \vec r)
  \partial_{\vec n}\phi(\vec s)-\phi(\vec s)\partial_{\vec n}g(\vec s, \vec r)
  \right\}ds &=& \phi^{\text{inc}}(\vec r), \quad \vec r\in \Gamma_a\,,
\end{eqnarray}
\begin{eqnarray}\label{eq:condition_virtualboundary}
-\int_{\partial\Omega_a}\left\{g(\vec s, \vec r)\partial_{\vec n}\phi(\vec s)
-\phi(\vec s)\partial_{\vec n}g(\vec s, \vec r)\right\}ds &=& \phi^{\text{sc}}(\vec r),
\quad \vec r\in \Gamma_b\,,
\end{eqnarray}
\begin{eqnarray}
\label{eq:condition_quietzone}
-\int_{\partial\Omega_a}\left\{g(\vec s, \vec r)\partial_{\vec n}\phi(\vec s)
-\phi(\vec s)\partial_{\vec n}g(\vec s, \vec r)\right\}ds &=& -\phi^{\text{inc}}
(\vec r), \quad \vec r\in \Gamma_c\,.
\end{eqnarray}
\end{subequations}
\end{widetext}
The solution to these integral equations can be numerically determined using
the boundary element method (BEM),~\cite{Xiao08} which is based on the SIE.
BEM approximates the surface integrals by discretizing the surface
$\Gamma_a\equiv\partial \Omega_a$ into $N$ surface elements $\ell _\alpha $ on
which the functions $\phi ({\rm {\bf s}})$ and $\psi ({\rm {\bf
s}})=\partial _{\rm {\bf n}} \phi ({\rm {\bf s}})$ are approximated
as constants that represent the value of functions $\phi ({\rm {\bf
s}})$ and $\psi ({\rm {\bf s}})$ across the entire element $\ell
_\alpha $, respectively. In other words, a local step function basis
with regard to $\ell _\alpha $ is used to expand $\phi ({\rm {\bf
s}})$ and $\psi ({\rm {\bf s}})$ over the entire surface $\partial
\Omega $, with the expansion coefficients denoted as $\phi _\alpha $
and $\psi _\alpha $, where $\alpha =1,2,\cdot \cdot \cdot N$. From
the perspective of BEM, these expansion coefficients can be viewed
as $2N$ active sources to be determined, and
Eq.~\eqref{eq:conditions_total} represents the conditions to determine
these $2N$ unknowns. Equation~\eqref{eq:condition_self} gives $N$ constraints.
Besides these, one can choose $N_c$ sample points
on $\Gamma _c$ and $N_b $ points on $\Gamma _b $ and then get a total of $N_c
+N_b + N$ constraints to determine the $2N$ degree of freedom $\phi _\alpha$ and
$\psi _\alpha$. We can then
establish the following linear equations
\begin{eqnarray}\label{eq:linear_equations}
\label{eq:asymble2} \left[
\begin{array}{l l}
{\vec H}_{ca} &{\vec G}_{ca}\\
{\vec H}_{ba} & {\vec G}_{ba}\\
{\vec H}_{aa}^{\text{int}} & {\vec G}_{aa}^{\text{int}}
\end{array}\right]\left[
\begin{array}{c c}
{\vec \Phi_a} \\ {\vec \Psi_a}
\end{array}\right]=
\left[
\begin{array}{r}
 {-\vec \Phi}^{\text{inc}}_c \\{\vec \Phi}^{\text {sc}}_b \\{\vec \Phi}^{\text{inc}}_a
\end{array}\right],
\end{eqnarray}
where ${\rm {\bf H}}_{ba} $, ${\rm {\bf G}}_{ba} $ and ${\rm {\bf
H}}_{ca} $, ${\rm {\bf G}}_{ca} $ represent the interacting matrices
relating the ``source points" on $\Gamma _a $ to the field points on
$\Gamma _b $ and $\Gamma _c $, and have elements defined as
\begin{subequations}
\begin{eqnarray}
 H_{\alpha\beta}&=&\int_{\ell_{\beta}}
\partial_{\vec n} g({\vec s}, {\vec
r_\alpha})ds\,, \\
G_{\alpha\beta}&=&-\int_{\ell_{\beta}} g( {\vec s}, {\vec r_\alpha})ds \\
&~&{\rm {\bf r}}_\alpha \in \Gamma _c \cup
\Gamma _b , \alpha=1, 2, ..., N_c+N_b, \nonumber\\
&~&\ell _\beta \subset \Gamma _a, \beta = 1, 2, ..., N\,.\nonumber
\end{eqnarray}
\end{subequations}
whereas, ${\rm {\bf H}}_{aa}^{\text{int}}$ and ${\rm {\bf G}}_{aa}^{\text{int}}$
represent the self-consistent conditions imposed on $\phi$ and $\partial_{\vec n}\phi$,
with elements defined as,
\begin{subequations}
\begin{eqnarray}
 H_{\alpha\beta}^{\text{int}}&=&\frac{1}{2}\delta_{\alpha\beta}-\int_{\ell_{\beta}}
\partial_{\vec n} g({\vec s}, {\vec r_\alpha})ds\,, \\
G_{\alpha\beta}^{\text{int}}&=&\int_{\ell_{\beta}} g({\vec s}, {\vec r_\alpha} )ds \\
&~&\ell_\alpha ,\ell _\beta \subset \Gamma _a, \alpha, \beta = 1, 2, ..., N,\nonumber\\
&~&{\rm {\bf r}}_\alpha
\text{~is the center of~} \ell_\alpha \nonumber \,.
\end{eqnarray}
\end{subequations}
On the right hand side of Eq.~\eqref{eq:asymble2}, ${\vec \Phi}^{\text{inc}}_c$  denotes the
incoming probing wave fields $\phi ^{\text{inc}}$ sampled at the inner quite
zone boundary $\Gamma _c \equiv \partial \Omega _c $, ${\vec
\Phi}^{\text{sc}}_{b}$ denotes the sampled scattered fields on the outer
boundary $\Gamma _b \equiv
\partial \Omega _b $ that would have been scattered by the object $V_i $
under the illumination of the same probing wave $\phi ^{\text{inc}}$.
${\vec \Phi}^{\text{inc}}_a$ denotes the incoming wave fields on the boundary
of the cloaking devices.
Thus, the fields outside $\Gamma _b $ (the dashed-line in
Fig.~\ref{fig1}) approach those of $\phi ^{\text{inc}}$ scattered
by the object $V_i$ as the number of sampling points(N) increases and the discrepancy decreases if N increases.  
If we set ${\vec \Phi} _b^{\text{sc}}=0$, we achieve the active external
invisible cloaking. Cloaking is thus a special case of illusion in this formulation. $\vec
\Phi_a$ and $\vec \Psi_a$ are both $N$-dimensional vectors representing the
total field and field gradient on $\Gamma_a$, which require active sources
to generate. For simplicity, in our numerical calculations, we set $N_c + N_b + N= 2N$, so that the matrix in Eq~\eqref{eq:linear_equations} is a square matrix. The linear system of equations is solved using the LAPACK subroutine ZGESV. We can see that BEM offers a physically transparent way of deriving the active sources needed to do remote cloaking and illusion and offers a straight forward numerical recipe in calculating those sources.

Our forgoing discussions are restricted to non-radiating objects. If the object inside $\Omega_c$ itself is a radiating source, we should add extra active sources to cancel the radiated field $\phi^{\text{radiate}}$ outside $\Gamma _b$. This extra term is determined by
\begin{widetext}
\begin{subequations}\label{eq:conditions_extra}
\begin{eqnarray}\label{eq:condition_self_extra}
  \frac{1}{2}\phi^{\text{extra}}(\vec r) + \int_{\partial\Omega_a}\left\{\tilde{g}(\vec s, \vec r)
  \partial_{\vec n}\phi^{\text{extra}}(\vec s)-\phi^{\text{extra}}(\vec s)\partial_{\vec n}\tilde{g}(\vec s, \vec r)
  \right\}ds &=& \phi^{\text{radiate}}(\vec r), \quad \vec r\in \Gamma_a\,,
\end{eqnarray}
\begin{eqnarray}\label{eq:condition_virtualboundary_extra}
-\int_{\partial\Omega_a}\left\{\tilde{g}(\vec s, \vec r)\partial_{\vec n}\phi^{\text{extra}}(\vec s)
-\phi^{\text{extra}}(\vec s)\partial_{\vec n}\tilde{g}(\vec s, \vec r)\right\}ds &=&-\phi^{\text{radiate}}(\vec r),
\quad \vec r\in \Gamma_b\,,
\end{eqnarray}
\begin{eqnarray}
\label{eq:condition_quietzone_extra}
-\int_{\partial\Omega_a}\left\{\tilde{g}(\vec s, \vec r)\partial_{\vec n}\phi^{\text{extra}}(\vec s)
-\phi^{\text{extra}}(\vec s)\partial_{\vec n}\tilde{g}(\vec s, \vec r)\right\}ds &=& 0, \quad \vec r\in \Gamma_c\,.
\end{eqnarray}
\end{subequations}
\end{widetext}
The Green's function $\tilde{g}$ might be different since the radiating field may be of a different frequency $\tilde{\omega}$. Then the total active field should be
\begin{eqnarray}
   \phi_a(\vec r, t)=\phi(\vec r)\exp{[-i\omega t]} + \phi^{\text{extra}}(\vec r)\exp{[-i\tilde{\omega} t]}\,,
\end{eqnarray}
where $\phi(\vec r)$ is the solution of Eq.~\eqref{eq:conditions_total}, and $\phi^{\text{extra}}$ is the solution of Eq.~\eqref{eq:conditions_extra}. We note that the condition in Eq.\eqref{eq:condition_quietzone_extra} is necessary since the radiating object may also be a passive scatterer. 
\section{Numerical simulations}
\subsection{Remote Cloaking Effect}
\label{subsec:cloaking}
\begin{figure*}[hbt]
\centering
\includegraphics[width=0.80\textwidth]{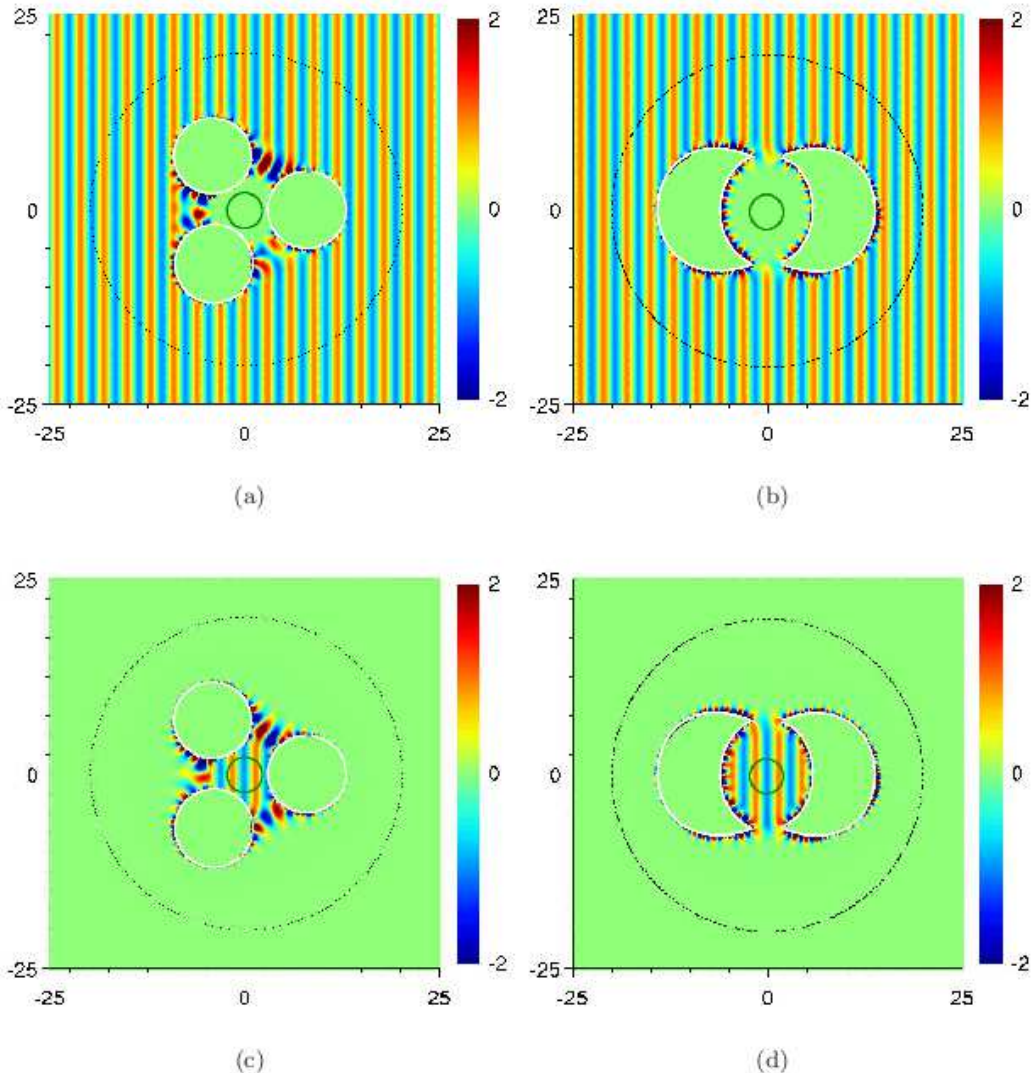}
\caption{(Color online) Example of active source external
cloak. (a),(b) show the total fields, and (c),(d) show the
``scattered" fields. The sources are arranged on the boundaries (white solid line)
of three circles in the left panels while are in two
crescent-shaped curves in the right panels. Here, the incoming plane wave is of wavelength $\lambda=3.0$. }\label{fig2}
\end{figure*}

Here, we show numerically that external cloaking is achieved by simply setting $\vec
\Phi_b^{\text{sc}}=0$ in Eq.~\eqref{eq:asymble2}. The choice of plane wave
$\exp(i\vec k\cdot \vec r)$ as the incoming source is just for simplicity but
the formulation works for other forms of incident wave. 
The configurations of the cloaking devices are shown in
Fig.~\ref{fig2} for two kinds of source arrangement. In the left panels, active
sources are placed on three circles arranged as shown in Fig.~\ref{fig2}(a) and
Fig.~\ref{fig2}(c). In the right panels, we show a case in which the active 
sources are placed on two crescent-shaped curves [Fig.~\ref{fig2}(b) and Fig.~\ref{fig2}(d)]. In both
cases, we have chosen $N=900$, $N_c=300$, and $N_b=600$. 
Employing the
scheme described in the preceding section, we can achieve an
approximate solution numerically. We see from Fig.~\ref{fig2} that the field inside the 
quiet zone is essentially zero and here is no scattering. The numerical solutions only
ensure that Eq.~\eqref{eq:conditions_total} is correct in a finite number of points. 
To quantify the overall quality of the solution, we consider the following error functions
 (measured with $L^2$ norm) defined on the two circles $\Gamma_b$ (r=20) and $\Gamma_c$ (r=2) and
 and inside the quiet zone $\Omega_c$, 
\begin{subequations}\label{eq:error_measure}
\begin{eqnarray}
  \text{Err}(\Gamma_b) = \frac{\oint_{\partial\Omega_b} |\phi(\vec s) -
  \phi^{\text{inc}}(\vec s)|^2 ds}{\oint_{\partial\Omega_b} |\phi^{\text{inc}}
  (\vec s)|^2 ds}\,,
\end{eqnarray}
\begin{eqnarray}
  \text{Err}(\Gamma_c) = \frac{\oint_{\partial\Omega_c} |\phi(\vec s)|^2 ds}
  {\oint_{\partial\Omega_c} |\phi^{\text{inc}}(\vec s)|^2 ds}\,,
\end{eqnarray}
\begin{eqnarray}
  \text{Err}(\Omega_c) = \frac{\iint_{\Omega_c} |\phi(\vec r)|^2 dA}
  {\iint_{\Omega_c} |\phi^{\text{inc}}(\vec r)|^2 dA}\,.
\end{eqnarray}
\end{subequations}
Each integrand is numerically evaluated at a set of 40,000 points sampled in 
the corresponding integration domain. For fixed circle sizes of $\Gamma_b$ and $\Gamma_c$, the errors depend on the choice of $N$, $N_b$ and $N_c$, as well as the frequency of the
incoming wave. A detailed discussion on the dependence of these errors on the parameters
can be found in Appendix~\ref{Sec:Error}. It is seen that the error decreases
as we increase N. In other words, we are able to achieve better cloaking
effects if we can control the boundary fields more precisely. The field
patterns are presented in Fig.~\ref{fig2}. Figs.~\ref{fig2}(a) and
~\ref{fig2}(b) show the total fields which are the superposition of the
incoming plane wave and the active fields generated
 by the ``active sources" [solutions to Eq.~\eqref{eq:asymble2}]
 placed on the white dotted-lines. The total fields $\phi ({\rm
{\bf r}})$ outside $\Gamma_b$ (marked by the black dashed-line), a circle of
$r=20$ units, are almost the same as the incoming plane wave, with a
discrepancy $\text{Err}(\Gamma_b)=4.62\times 10^{-13}$ in Figs.~\ref{fig2}(a)
and \ref{fig2}(c). At the same time, we achieve a ``quiet" zone $\Omega_c$
(bounded by $\Gamma_c$ as marked by the green lines) within a circle of $r=2$
inside which the total fields almost vanish, with
$\text{Err}(\Gamma_c)=1.14\times 10^{-12}$ and
$\text{Err}(\Omega_c)=1.03\times 10^{-12}$. Figs.~\ref{fig2}(c) and
\ref{fig2}(d) show the corresponding ``scattered'' fields $\phi
^{\text{sc}}({\rm {\bf r}})=\phi ({\rm {\bf r}})-\phi^{\text{inc}}({\rm {\bf
r}})$, which are exactly the fields
 created by the active devices. Concomitantly, $\phi
^{\text{sc}}({\rm {\bf r}})$ vanishes outside $\Gamma _b $ and is the reverse
of $\phi ^{\text{inc}}({\rm {\bf r}})$ inside the ``quiet" zone $\Omega _c $.
The strength of the fields on the boundary $\Gamma_a$ is in the order of 100
[e.g., $\sim160$ in Figs.~\ref{fig2}(a) and (c), $\sim80$ in
Fig.~\ref{fig2}(b) and (d)], and can be easily achieved physically. Vasquez and co-workers~\cite{MiltonPRL} 
proposed that three disjoint circular disks are needed to perform remote active cloaking. Here, we see that 
from Figs.~\ref{fig2}(b) and ~\ref{fig2}(d) that a non-circular cloaking device
comprising two simply connected regions can also achieve the similar cloaking effect.
It is further shown (figure not presented here) that active sources on one simply connected 
cloaking device can also achieve a high degree of invisibility.

\subsection{Illusion Effect}
\label{subsec:illusion}
\begin{figure*}[hbt]
\centering
    \includegraphics[width=0.95\textwidth]{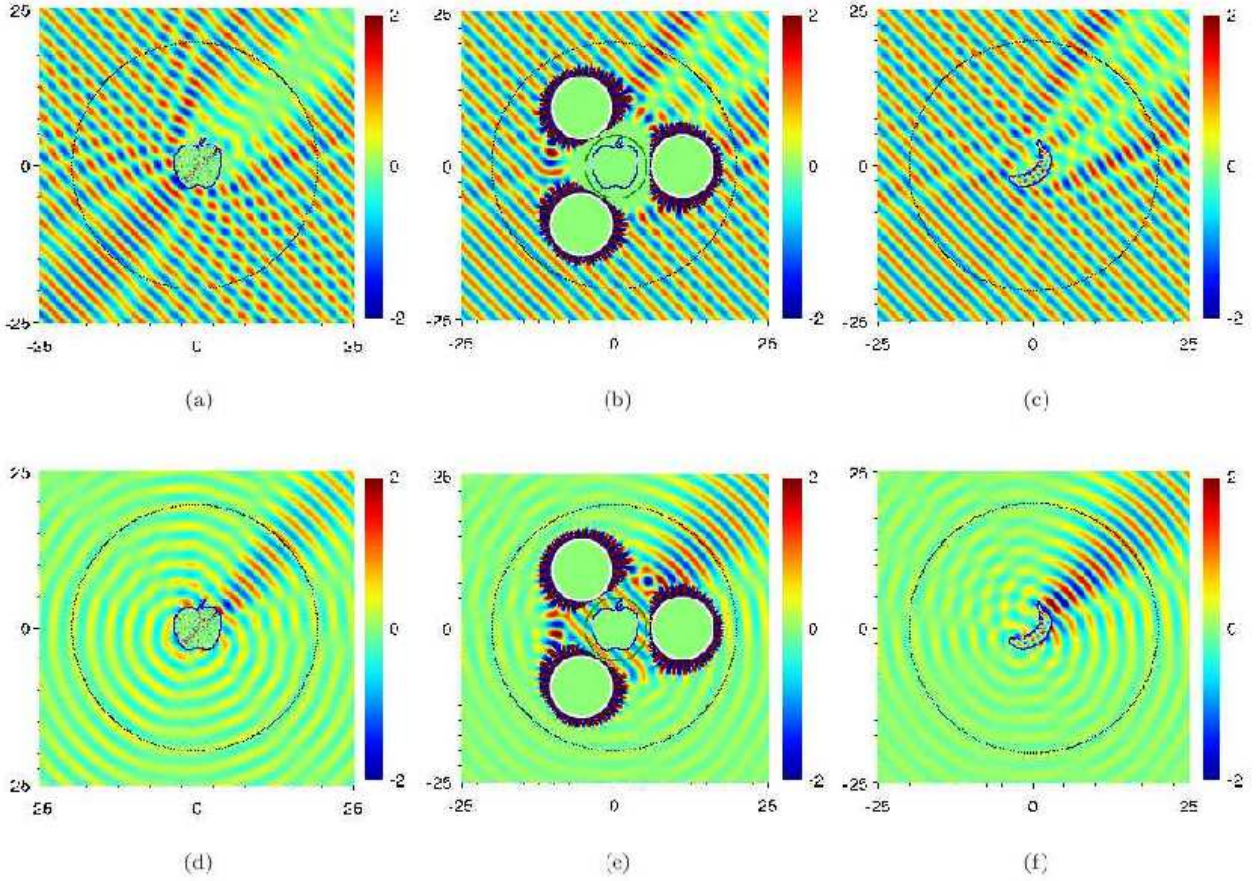}
\caption{(Color online) Optical illusion effect by active sources.
(a), (b) and (c) show the total fields. (d), (e) and (f) show the
``scattered'' fields ($\phi^{\text{tot}}-\phi^{\text{inc}}$). (a) and (d) are for an apple ($\epsilon=16.0, \mu=1.0$). (b) and (e) are for an illusion device with an apple-shaped object [identical to (a) and (d)] concealed inside $\Omega_c$. The active sources are placed on the three circles (white solid line). (c) and (f) are for a banana-shaped dielectric object ($\epsilon=5.0, \mu=1.0$). The active sources give the same
scattering pattern as the ``banana" outside the black dotted
curve. }\label{fig3}
\end{figure*}

\begin{figure*}[tbh]
\centering
\includegraphics[width=0.95\textwidth]{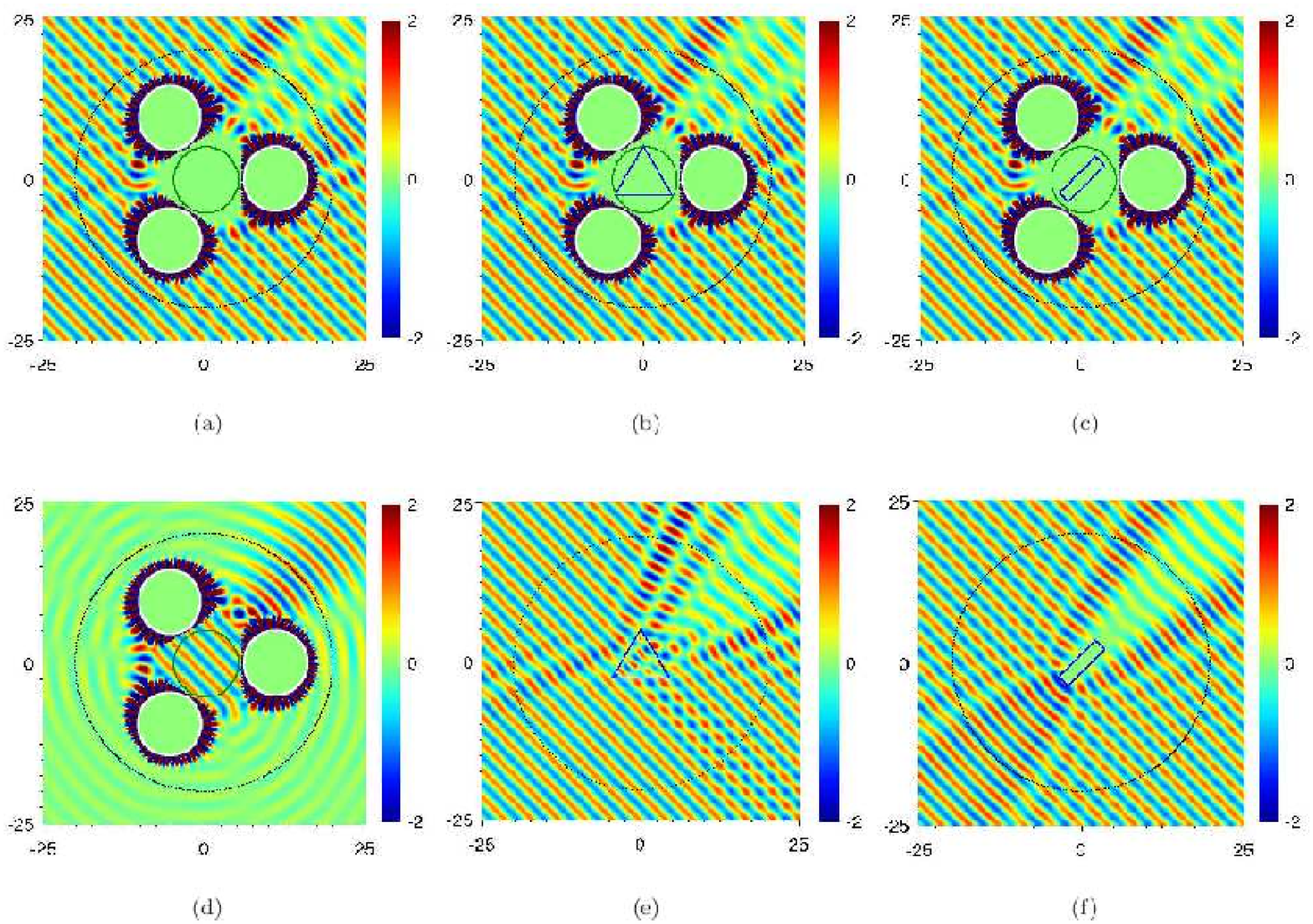}
\caption{(Color online) Optical illusion effect by active sources.
(a), (d) show the total field pattern of an illusion device. (b) and (c) are patterns of the total fields for the same illusion device but with different objects concealed inside $\Omega_c$ [(b) is a triangle $\epsilon=2.25,\mu=-1.0$, (c) is a PEC slab]. (e) and (f) are patterns of the total fields for the objects identical to those concealed inside the ``quiet" zones in (b) and (c). Any object can be concealed inside the ``quiet'' zone where the total fields are almost zero.}\label{fig4}
\end{figure*}

Next, we demonstrate an illusion effect such that whatever
objects placed inside the ``quiet'' zone $\Omega_c$ will appear like
another object, which is chosen here to be a banana-shaped dielectric object with refractive index $n=2.32$. Figs.~\ref{fig3}(c) and
\ref{fig3}(f) show the total and scattered fields of such a ``banana" under
the illumination of a plane wave $\exp (i\vec k\cdot \vec r)$.  We 
set the control boundary $\Gamma _b$ (marked as the dashed-line) as a circle
with radius $r=20$ outside which the illusion shall be observed. This
requires one to set $\vec \Phi _b^{\text{sc}} $ in Eq.~\eqref{eq:asymble2} as
the scattered fields due to the pre-designed ``banana" [see Fig.~\ref{fig3}(c)] on
$\Gamma_b$. A total of $N=900$ sample points (with $N_b=600, N_c=300$) are used here in the numerical
calculation, and $\vec \Phi _c^{\text{inc}}$ in Eq.~\eqref{eq:asymble2} is
the function $\exp(i\vec k\cdot \vec r)$ sampled over a circle of $r=5$ (the
boundary of the ``quiet" zone $\Omega_c$ as marked by the green solid
circle). In Figs.~\ref{fig3}(b) and \ref{fig3}(e), we conceal
an apple-shaped dielectric object ($n=4$) inside the ``quiet" zone. Figs.~\ref{fig3}(a) and \ref{fig3}(d) 
show the total and scattered fields of the ``apple" under the illumination of the plane wave. We see in Figs.~\ref{fig3}(b) and \ref{fig3}(e) that, after turning on the active sources, the total system responses to the incoming plane wave in a way as if a ``banana" is placed in the cloaked region
[Figs.~\ref{fig3}(c) and \ref{fig3}(f)]. The field discrepancy on the circle of $r=20$ is
$1.17\times 10^{-6}$, measured with $L^2$ norm as,
\begin{eqnarray}
  \text{Err}^{ill}(\Gamma_b) = \frac{\oint_{\partial\Omega_b} |\phi(\vec s)
  - \phi^{\text{inc}}(\vec s)-\tilde{\phi}^{\text{sc}}(\vec s)|^2 ds}
  {\oint_{\partial\Omega_b} |\phi^{\text {inc}}(\vec s)|^2 ds}\,,
\end{eqnarray}
where $\tilde{\phi}^{\text{sc}}$ is the scattered field of the pre-designed
``banana" under the illumination of the incoming wave.
Fig.~\ref{fig3}(e) shows that the active sources construct a negative
counterpart of the incoming plane wave to create the ``quiet'' zone $\Omega_c$, so that the total field surrounding the ``apple" is almost zero. We note that we can put any non-radiating object inside
$\Omega_c$ without affecting the total scattering pattern outside $\Gamma_b$
[see Figs.~\ref{fig4}(b) and \ref{fig4}(c)], since all the fields inside $\Omega_c$ are almost
zero as is shown in Fig.~\ref{fig4}(a) [The discrepancies: 
$\text{Err}^{ill}(\Gamma_c)=6.15\times 10^{-7}$, 
$\text{Err}^{ill}(\Omega_c)=1.11\times 10^{-6}$]. In fact, any passive
object inside $\Omega_c$ does not ``talk" to the other parts of the world, rendering
this illusion device workable for multiple and arbitrary objects. In
principle, by changing the active sources, we can let observers outside
$\Gamma_b$ see whatever we want them to see. 

\subsection{Cloaking a radiating object}
\begin{figure*}[hbt]
\centering
   \includegraphics[width=0.95\textwidth]{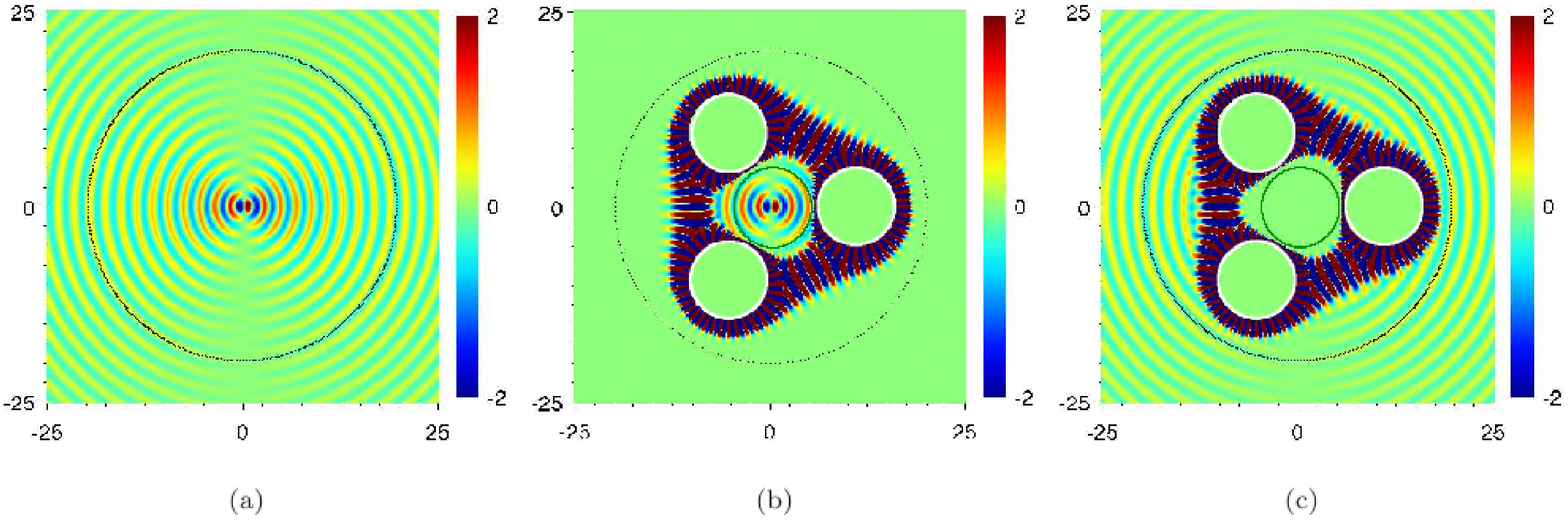}
\caption{(Color online) Cloaking a radiating object.
(a)The radiating field pattern of $10\cdot H_1^{(1)}(\tilde{k}r)\cos\theta\exp[-i\tilde{\omega} t]$, where $\tilde{k}=\pi$. (b) The total field pattern with the sources on the boundary of the cloaking device actives(three circles) radiating according to the solutions of Eq.~\eqref{eq:conditions_extra}. (c) The field generated by the cloaking device.}\label{fig5}
\end{figure*}

Here, we demonstrate the cloaking effect of a radiating object. For simplicity, we assume that $\phi^{\text{radiate}}=10\cdot H_1^{(1)}(\tilde{k}r)\cos\theta \exp[-i\tilde{\omega} t]$ and there is no incoming wave from the outside. The simulation results are shown in Fig.~\ref{fig5}, with $\tilde{k}=\pi$. Here, we use the same parameters $N, N_b, N_c$ as those in section~\ref{subsec:cloaking} and section~\ref{subsec:illusion}. From  Fig.~\ref{fig5}(b), we see that the total field pattern near the dipole remain the same with Fig.~\ref{fig5}(a). The cloaking device creates active fields which cancel the radiating fields outside $\Gamma_b$ [see Figs.~\ref{fig5}(b) and \ref{fig5}(c)] and do not affect the field pattern inside $\Omega_c$. The active source in this example is just the $\phi^{\text{extra}}(\vec s)$ which is the solution of Eq.~\eqref{eq:conditions_extra}. Considering the linear superposition property, one can add together the solution of Eq.~\eqref{eq:conditions_total} and $\phi^{\text{extra}}$ to achieve cloaking or illusion effects for this specific radiating object. 
\section{Conclusion}
In conclusion, we applied a boundary element method to demonstrate both external cloaking
and illusion effects using active sources on continuous curves for the 2D Helmholtz equation.
The scalar wave formalism applies to both acoustic waves and electromagnetic waves in two dimensions.
This approach works for arbitrary objects and there is no intrinsic bandwidth
limitation. The limitation of this type of remote active cloaking and illusion is that it requires the 
prior knowledge of the incoming wave or the availability of sensors that can detect the fields quickly
enough on the boundaries and of active sources that can respond fast enough.~\cite{Miller06}

\hfill

\acknowledgments{This work was supported by Hong Kong RGC Grant
No.~600209. Computation resources are supported by Shun Hing
Education and Charity Fund. We acknowledge discussions with Z. H.
Hang and Jeffery Lee.}

\begin{appendix}
\section{Possible ways to build up the active sources}
The boundary element method gives the field and field gradients on the boundary of the 
active cloaking device. It does not give directly the sources that give rise to those 
fields. There are obviously many different ways of arranging sources within the boundary 
of the active devices to generate the necessary fields. In this Appendix, we give a few 
examples. 
\subsection{Using monople sources and dipolar sources located on the boundaries}
\begin{figure}[hbt]
   \centering
   \includegraphics[width=0.40\textwidth,clip]{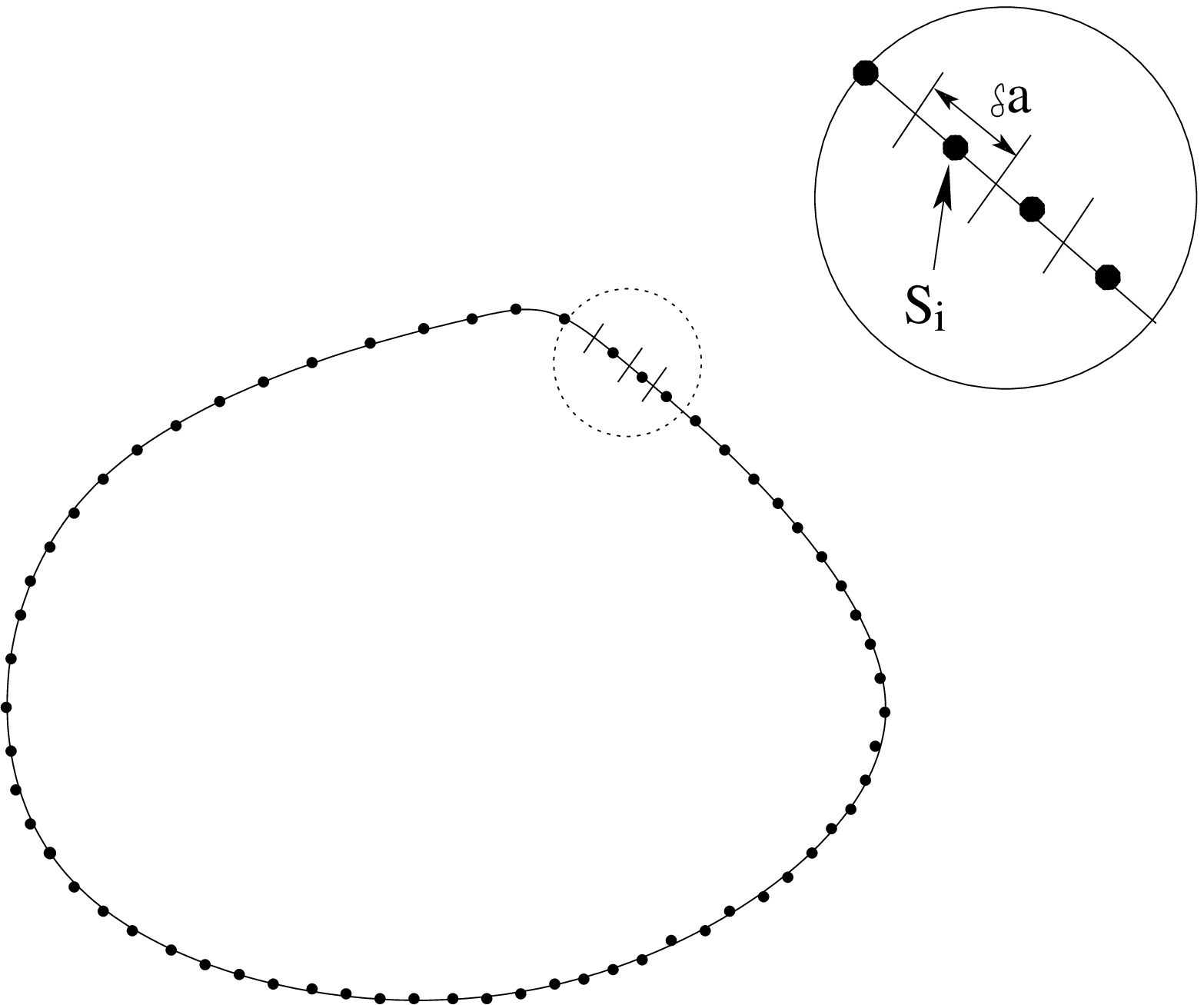}
   \label{fig:singlelayersources}
   \caption{Generating active field by using monople sources and dipolar sources located on the boundaries of the cloaking or illusion devices.}
\end{figure}
We note that the active fields are given by,
\begin{equation}\label{eq:activemethod1appendix}
    \phi^{\text{active}}(\vec r)=-\oint_{\partial \Omega_a}ds
    \big[g({\vec s}, {\vec r}) \partial_{{\vec n}}\phi_a({\vec s})-\phi_a({\vec s})
    \partial_{\vec{n}} g({\vec s},{\vec r})\big]\,,
\end{equation}
where $\phi_a$ and $\partial_n\phi_a$ are the solutions of Eq.~\eqref{eq:conditions_total}. The first term represents a set of monopole point sources, with the corresponding strengths as follows 
\begin{eqnarray}\label{eq:singlelayermonopole}
p_i = -\partial_n\phi_a(\vec s_i)\delta a_i\,,
\end{eqnarray}
where $\vec s_i$ is the center of the i-th boundary element, and $\delta a_i$ is the length of the corresponding element. The second term represents a set of dipole sources on the boundary, since
\begin{eqnarray}
   \partial_n g(\vec s, \vec r) &=& \frac{ik}{4}H_{1}^{(1)}(k|\vec r-\vec s|)\frac{\vec r-\vec s}{|\vec r-\vec s|}\cdot \vec n \nonumber \\&=&  \frac{ike^{i\theta'}}{8}H_{1}^{(1)}(k|\vec r-\vec s|)e^{i\theta}\nonumber\\
   &~&+\frac{ike^{-i\theta'}}{8}H_{1}^{(1)}(k|\vec r-\vec s|)e^{-i\theta}\,,
\end{eqnarray}
which are exactly dipole sources located at $s_i$. Here, $\theta'$ is the direction of the unit normal $\vec n$ and $\theta$ is the direction of $\vec r-\vec s$. We have used the recursion relation of Hankel functions $\frac{dH_{0}^{(1)}(x)}{dx} = - H_{1}^{(1)}(x)$.

The corresponding strengths of the two types of dipole sources are 
\begin{eqnarray}\label{eq:singlelayerdipole}
q_{i_1} &=& \frac{ike^{i\theta'_i}\delta a}{8}\phi_a(\vec s_i), \quad \text{ for } H_{1}^{(1)}(k|\vec r-\vec s_i|)e^{i\theta}\,,\nonumber\\
q_{i_2} &=& \frac{ike^{-i\theta'_i}\delta a}{8}\phi_a(\vec s_i), \quad \text{ for } H_{1}^{(1)}(k|\vec r-\vec s_i|)e^{-i\theta}\,.
\end{eqnarray}
It can be easily verified that this set of point sources and dipole sources can exactly generate the necessary active fields [see the schematic configuration on Fig.~\ref{fig:singlelayersources}]. 

\subsection{Using double layers of point sources}
\begin{figure}[hbt]
   \centering
   \includegraphics[width=0.45\textwidth,clip]{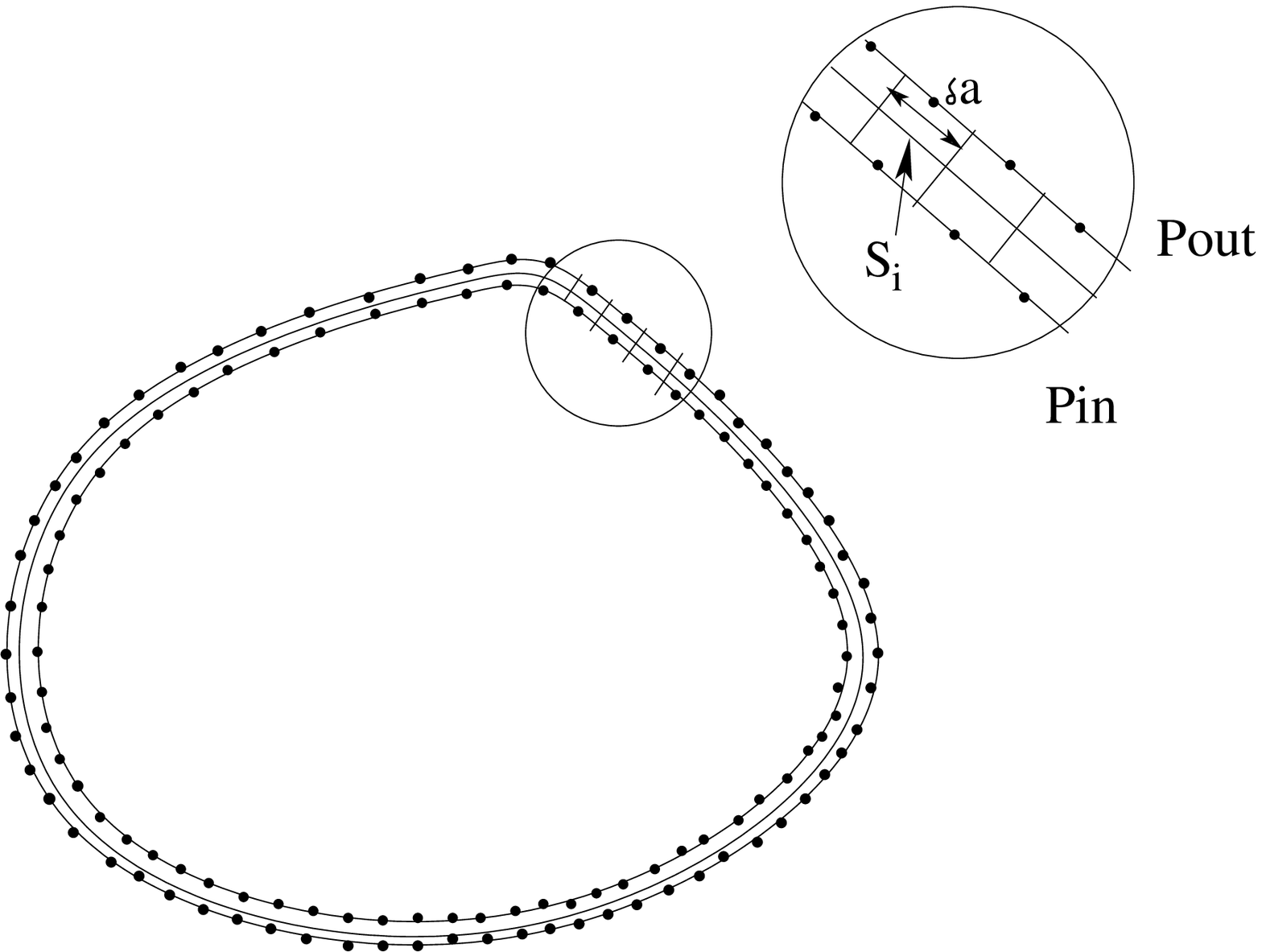}
   \caption{Generating active field by using point sources located on double layer boundaries.}
   \label{fig:doublelayersources}
\end{figure}

We can also use only point sources to build up the same necessary active fields ~\cite{Miller06}. Instead of positioning the sources on the boundary $\Gamma_a$, one can place point sources on the outer boundary $S_{out}$ and the inner boundary $S_{in}$ [see the schematic configuration on Fig. ~\ref{fig:doublelayersources}], i.e. we place point sources at the following points
\begin{eqnarray}
\vec s^{out}_i = \vec s_i + \frac{h}{2}\vec n_i, \quad
\vec s^{in}_i = \vec s_i - \frac{h}{2}\vec n_i,\quad i=1, 2, ..., N\,,
\end{eqnarray}
where $\vec s_i$ is the center of the i-th element and $\vec n_i$ is the unit normal of the i-th element. 
The corresponding strengths of the point sources are, 
\begin{eqnarray}
  p(\vec s^{out}_i) &=& \frac{\delta a_i}{h}\Big[\phi_a(\vec s_i)-\frac{h}{2}\partial_n\phi_a(\vec s_i)\Big],\nonumber\\
  p(\vec s^{in}_i) &=& -\frac{\delta a_i}{h} \Big[\phi_a(\vec s_i)+\frac{h}{2}\partial_n\phi_a(\vec s_i)\Big]\,.
\end{eqnarray}
It can easily seen that the field generated by the point sources above is
\begin{eqnarray}
  \phi(\vec r)&=&\sum_{i}^{N}\Big[p(\vec s^{out}_i)g(\vec s^{out}_i, \vec r) + p(\vec s^{in}_i)g(\vec s^{in}_i, \vec r)\Big]\nonumber\\
  &=&-\oint_{\partial \Omega_a}ds
    \big[g({\vec s}, {\vec r}) \partial_{{\vec n}}\phi_a({\vec s})-\phi_a({\vec s})
    \partial_{\vec{n}} g({\vec s},{\vec r})\big]\nonumber\\
    &~&+O(h^2)\,.
\end{eqnarray}
As long as the distance h between the outer and the inner boundary is sufficiently small, one can build up the active field as close as we like to that of Eq.~\eqref{eq:activemethod1appendix}. 
\subsection{Using multipole sources placed at the centers of the cloaking or illusion devices}
\label{Sec:Source} 
Instead of positioning monopole or dipole sources directly on the boundaries of the cloaking or illusion devices, one can put multipole active sources inside the cloaking devices
$\Omega_a$ to construct the required boundary fields $\phi$ and
$\partial_{\vec n}\phi$, and then control the fields outside $\Omega_a$. For a circular domain, it is convenient to place the multipole source at the center. In the following, we will discuss a way to determine the corresponding strength of each component from the boundary fields
obtained by BEM.

We note that the active fields are given by,
\begin{equation}\label{eq:activeappendix}
    \phi^{\text{active}}(\vec r)=-\oint_{\partial \Omega_a}ds
    \big[g({\vec s}, {\vec r}) \partial_{{\vec n}}\phi_a({\vec s})-\phi_a({\vec s})
    \partial_{\vec{n}} g({\vec s},{\vec r})\big]\,.
\end{equation}
For the domain outside the cloaking device, $|\vec s-\vec r_c|<|\vec r-\vec r_c|$
($\vec r_c$ is the center of the cloaking device), we can expand the Green's function
as,
\begin{eqnarray}\label{eq:expansion}
  g(\vec r, \vec s) &=& \sum_{m=-\infty}^{\infty}H_m^{(1)}(k|\vec r-\vec r_c|)
  J_m(k|\vec s-\vec r_c|)\exp[im(\Delta\theta)], \nonumber\\
  &~&\quad |\vec s-\vec r_c|<|\vec r-\vec r_c|,
\end{eqnarray}
where $H_m^{(1)}$ is the $m$-th order Hankel function of the first type and $\Delta\theta$ is the angle
between $\vec r-\vec r_c$ and $\vec s-\vec r_c$.
Substituting this into Eq.~\eqref{eq:activeappendix}, one can get the
corresponding multipole sources. For example, for the case shown in
Fig.~\ref{fig2}(a), one can use multipole sources locating at the centers of the three circles
of the circles $r_{c_i}, i=1, 2, 3$ to generate the required fields and field gradients, i.e,
\begin{eqnarray}
  \phi^{\text{active}}(\vec r)=\sum_{i=1}^{3}\sum_{m=-\infty}^{\infty}a_{m, i}
  H^{(1)}_{m}(k|\vec r-\vec r_{c_i}|)\exp[im\theta]\,,\nonumber\\
\end{eqnarray}
where  $\theta$ is the direction of $\vec r-\vec r_{c_i}$, and $a_{m, i}$ is the strength of the multipole source $H^{(1)}_m$ locating at
$r_{c_i}$, which can be obtained by the following integrals,
\begin{eqnarray}\label{eq:ami1}
  a_{m, i}&=&\oint_{\partial\Omega_{a_{i}}}ds\exp[im\theta']\Big\{\partial_{\bf n}J_m(k|\vec s-
  \vec r_{c_i}|)~\phi_a(\vec s)\nonumber\\
  &~&-J_{m}(k|\vec s-\vec r_{c_i}|)~\partial_{\vec n}\phi_a(\vec s)\Big\}\,,
\end{eqnarray}
where $i=1, 2, 3$, $J_m(x)$ is the $m$-th order Bessel function and $\theta'$ is the direction of $\vec s - \vec r_{c_i}$.

The multipole expansion of the Green's function demonstrates that the two
kinds of boundary fields and their gradients ($\phi(\vec s)$ and $\partial_{\vec n}\phi(\vec s)$)
contribute the same form of multipole radiation fields $H_m(kr)\exp(im\theta)$, with their strengths
determined by the surface integral. In this sense, $\phi(\vec s)$ and $\partial_{\vec n}\phi(\vec s)$ are on the same footing in generating the active field in the open domain(i.e. $R^2-\Omega_a$).
This indicates that, in Eq.~\eqref{eq:activeappendix}, there are multiple choices of the
boundary fields $\phi_a$ and $\partial_{\vec n}\phi_a$, for the same $\phi^{\text{active}}(\vec r)$.
This is reasonable, as one can easily verify that,
\begin{eqnarray}\label{eq:zero}
    \oint_{\partial \Omega_a}ds
    \big[\phi^{k}({\vec s})
    \partial_{\vec{n}} g({\vec s},{\vec r})-g({\vec s}, {\vec r}) \partial_{{\vec n}}\phi^{k}({\vec s})\big]=0, \nonumber\\
    \quad\text{for}~\vec r \in R^2-\Omega_a
\end{eqnarray}
if $\phi^{k}(\vec r)$ satisfies the Helmholtz equation
\begin{eqnarray}
  (\nabla^2 + k^2)\phi^{k}(\vec r) = 0, \quad \vec r\in \Omega_a.
\end{eqnarray}
If a $\{\phi_a(\vec s), \partial_n\phi_a(\vec s)\}$ pair can give the required active sources, any solution
of the form $\{\phi_a(\vec s)+\phi^{k}(\vec s), \partial_{\vec n}\phi_a(\vec s) + \partial_{\vec n}
\phi^{k}(\vec s)\}$ gives the same active fields. Then a question arises:
Is the solution obtained by Eq.~\eqref{eq:asymble2} unique? In fact, we have imposed the continuity
boundary  condition across $\Gamma_a$,
\begin{eqnarray}
\lim_{\vec r\to \vec s}\phi(\vec r)=\phi(\vec s)|_{\vec s\in\partial\Omega_a}\,.
\end{eqnarray}
This automatically removes the extra degree of freedom, since any extra term $\phi^{k}(\vec s)$
makes the field discontinuous when taking the limit to the boundary. Physically,
if we are using multipole sources located inside the cloaking or illusion devices,
the continuity boundary conditions must be satisfied.

However, as we know from Eq.~\eqref{eq:zero}, the extra term $\phi^{k}$ does not
influence the active fields outside the cloaking device. Therefore, instead of
imposing the boundary continuity conditions, one can first find solutions of
the form $\{\tilde{\phi}_a, 0\}$ or $\{0, \partial_n\tilde{\phi}_a\}$, since
both $\phi$ and $\partial_\vec {n}\phi$ can independently provide the required
active sources ``mathematically''.  In fact, physically, $\{\tilde{\phi}_a, 0\}$ corresponds to placing dipoles on the boundaries and $\{0, \partial_n\tilde{\phi}_a\}$ corresponds to placing monopoles on the boundaries [see Eqs.~\eqref{eq:singlelayermonopole} and \eqref{eq:singlelayerdipole}]. It is seen that one can use only the monopoles or only the dipoles placed on the boundaries to obtain the required fields. For the $\{\tilde{\phi}_a, 0\}$ type of solution, the active fields can be expressed as,
\begin{eqnarray}
\phi^{{\text active }}(\vec r) = -\int_{\partial\Omega_a} g(\vec s, \vec r)\partial_{\bf n}\tilde{\phi}_a(\vec s)ds ,\quad \vec r\in R^2 -\Omega\nonumber\\
\end{eqnarray}
Under this circumstance, the conditions for cloaking that the active fields should satisfy are,
\begin{eqnarray}\label{eq:finaleq}
-\int_{\partial\Omega_a} g(\vec s, \vec r)\partial_{\bf n}\tilde{\phi}_a(\vec s)ds = \left\{
    \begin{array}{r l}
        -\phi^{{\text {inc}}}(\vec r), &\quad {\text {for}} ~\vec r\in \Gamma_c\,, \\
    0, &\quad {\text {for}} ~\vec r \in \Gamma_b\,.
    \end{array}\right.\nonumber\\
\end{eqnarray}
One can also find the corresponding multipole sources, with the following strength coefficients
\begin{eqnarray}\label{eq:ami2}
  a_{m, i}&=&-\oint_{\partial\Omega_{a_{i}}}ds ~\exp[im\theta']J_{m}(k|\vec s-\vec r_{c_i}|)~\partial_{\bf n}\tilde{\phi}(\vec s)\,.\nonumber\\
\end{eqnarray}
The set of $a_{m, i}$ should be the same with that obtained from the first
method in Eq.~\eqref{eq:ami1}. Please note that, in this case, we does not impose the boundary continuity conditions, therefore, if one takes the following
limit from outside the cloaking device,
\begin{eqnarray}\label{continuity}
\phi^{\text{phys.}}(\vec s)|_{\vec s\in\partial\Omega}&\dot{=}&\lim_{\vec r\to \vec s}\phi(\vec r)\,,\\
\partial_{\bf n}\phi^{\text{phys.}}(\vec s)|_{\vec s\in\partial\Omega} &\dot{=}&\lim_{\vec r \to\vec s}
\partial_{\bf n}\phi(\vec r)\,,
\end{eqnarray}
generally, $\phi^\text{phys.}(\vec s)$ is different with $\tilde{\phi}_a(\vec s)$. However, 
in principle, the fields $\{\phi^{\text{phys.}}, \partial_n\phi^{\text{phys.}}\}$ should be the same with the $\{\phi_a, \partial_n\phi_a\}$ acquired by imposing the boundary continuity conditions if the number of sample points N is large. 
This method can reduce the dimension of the matrix by half. 
We present the numerical simulation
results of different formalisms in Appendix~\ref{Sec:Error}.
\section{Convergence discussion}
\label{Sec:Error}

\begin{figure*}[tbh]
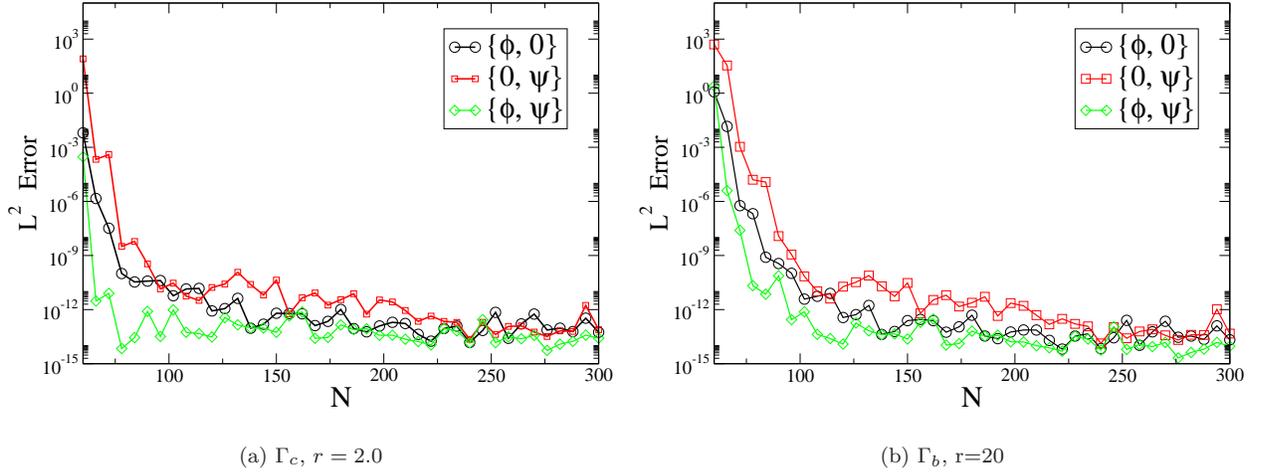

\centering
\subfigure[$\Gamma_c$, $r=2.0$]{
    \includegraphics[width=0.45\textwidth, clip]{delta-circ.eps}
    \label{fig:delta-circ-vnface}}
\subfigure[$\Gamma_b$, r=20]{
  \includegraphics[width=0.45\textwidth, clip]{gamma-circ.eps}
  \label{fig:gamma-circ-vnface}}
\caption{(Color online) Errors on (a) $\Gamma_c$: $r=2.0$ and $\Gamma_b$: $r=20.0$,
The wavelength $\lambda = 10.0$. The errors are measured with the $L^2$ norm, defined in Eq.\eqref{eq:error_measure}.}\label{error-vnface}
\label{Fig:Fig4}
\end{figure*}

\begin{figure*}[tbh]
\centering
    \includegraphics[width=0.90\textwidth, clip]{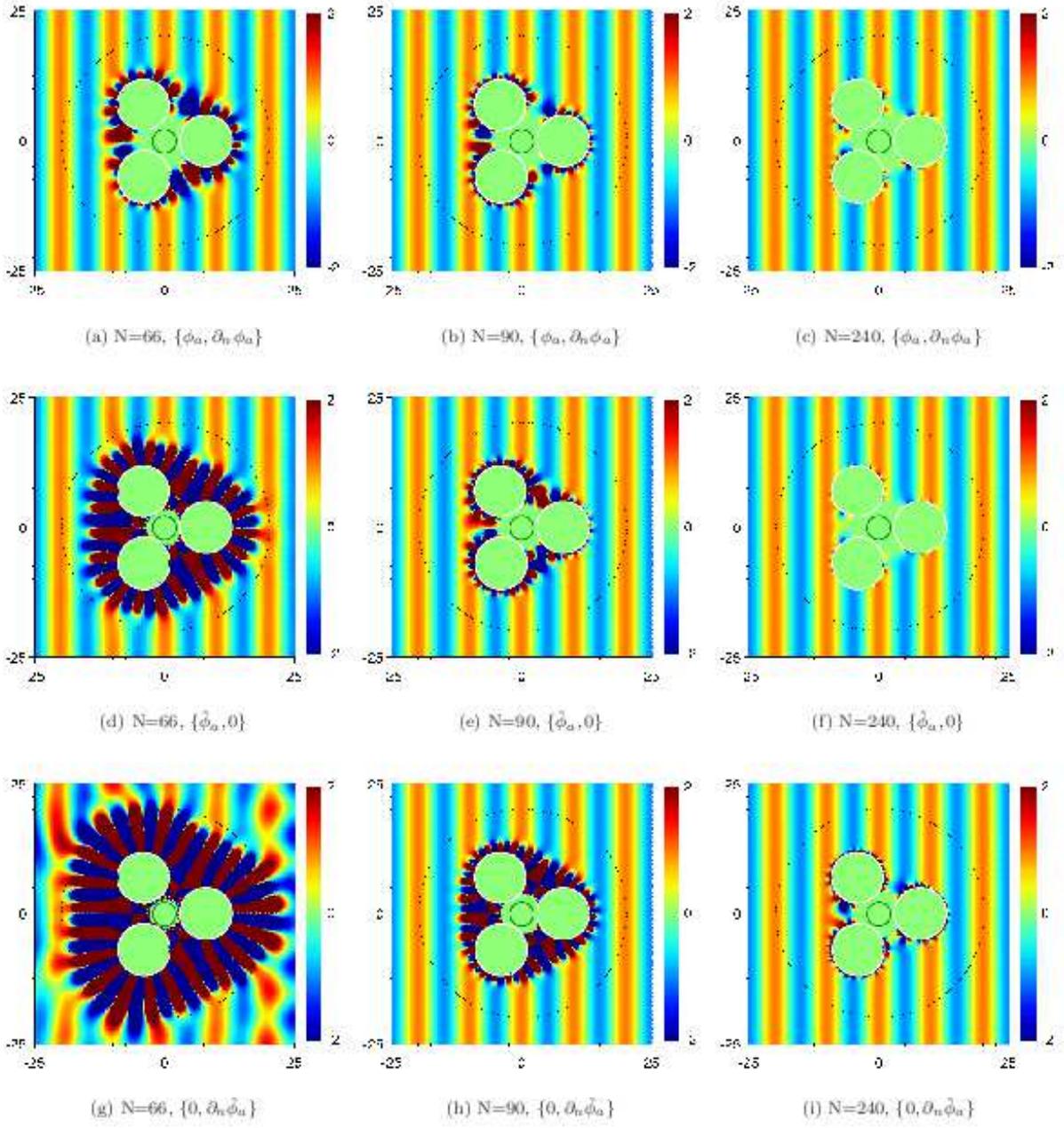}
\caption{(Color online) Field patterns of different schemes with different
$N$. $\Gamma_c: r=2.0$, $\Gamma_b: r=20.0, \lambda = 10.0$}
\label{field-pattern-compare}
\end{figure*}

The numerical solution of Eq.\eqref{eq:linear_equations} only ensure that Eq.~\eqref{eq:conditions_total} is satisfied for a finite number of points. Here, we consider the convergence of the solution as the number of sampling point increases by calculating the errors on the two circles
$\Gamma_c$ and $\Gamma_b$. The errors are measured as Eq.~\eqref{eq:error_measure} in the text. As shown in Fig.~\ref{Fig:Fig4},
the error decreases very quickly as $N$ increases (note that the vertical axis is in log scale). When $N$ is greater than 300 (this number, however,
depends on the wavelength and the geometry of the cloaking device), the total error is of $\sim10^{-14}$. This indicates that the field are essentially the same as the incoming wave outside the circle
of $r=20.0$ and nearly zero inside the quiet zone inside the circle of $r=2.0$, which confirms that
for all the practical
purposes, objects concealed inside the ``quiet'' zone are undetectable outside of $\Gamma_b$. To get
a better picture of the performance of the cloaking as $N$ increases, we provide some
representative field patterns of the three schemes in Fig. \ref{field-pattern-compare}. As can be clearly
seen, when $N$ is not large enough, there will be a lot of evident ``tails'' reaching out from the boundary of the cloaking device and the field values are large, while for increasing $N$,
the ``tails'' become less conspicuous. Meanwhile, the field patterns of the three
different schemes look indistinguishable with each other as $N$ increases.

\end{appendix}

\newpage

\begin{thebibliography}{23}
	 
\expandafter\ifx\csname
natexlab\endcsname\relax\def\natexlab#1{#1}\fi
\expandafter\ifx\csname bibnamefont\endcsname\relax
  \def\bibnamefont#1{#1}\fi
\expandafter\ifx\csname bibfnamefont\endcsname\relax
  \def\bibfnamefont#1{#1}\fi
\expandafter\ifx\csname citenamefont\endcsname\relax
  \def\citenamefont#1{#1}\fi
\expandafter\ifx\csname url\endcsname\relax
  \def\url#1{\texttt{#1}\fi
\expandafter\ifx\csname urlprefix\endcsname\relax\def\urlprefix{URL
}\fi \providecommand{\bibinfo}[2]{#2}
\providecommand{\eprint}[2][]{\url{#2}}

\bibitem[{\citenamefont{Kerker}(1975)}]{Kerker75}
\bibinfo{author}{\bibfnamefont{M.}~\bibnamefont{Kerker}}, \bibinfo{journal}{{J.
  Opt. Soc. Am.}} \textbf{\bibinfo{volume}{65}}, \bibinfo{pages}{376}
  (\bibinfo{year}{1975}).

\bibitem[{\citenamefont{Al\'{u} and Engheta}(2005)}]{Alu05}
\bibinfo{author}{\bibfnamefont{A.}~\bibnamefont{Al\'{u}}} \bibnamefont{and}
  \bibinfo{author}{\bibfnamefont{N.}~\bibnamefont{Engheta}},
  \bibinfo{journal}{{Phys. Rev. E}} \textbf{\bibinfo{volume}{72}},
  \bibinfo{pages}{016623} (\bibinfo{year}{2005}).

\bibitem[{\citenamefont{{L. S. Dolin}}(1961)}]{Dolin}
\bibinfo{author}{\bibnamefont{{L. S. Dolin}}}, \bibinfo{journal}{{Izv. Vyssh.
  Uchebn. Zaved. Radiofizika}} \textbf{\bibinfo{volume}{4}},
  \bibinfo{pages}{964} (\bibinfo{year}{1961}).

\bibitem[{\citenamefont{Leonhardt}(2006)}]{Leonhardt06}
\bibinfo{author}{\bibfnamefont{U.}~\bibnamefont{Leonhardt}},
  \bibinfo{journal}{{Science}} \textbf{\bibinfo{volume}{312}},
  \bibinfo{pages}{1777} (\bibinfo{year}{2006}).

\bibitem[{\citenamefont{Leonhardt and Philbin}(2006)}]{LeonhardtNJP}
\bibinfo{author}{\bibfnamefont{U.}~\bibnamefont{Leonhardt}} \bibnamefont{and}
  \bibinfo{author}{\bibfnamefont{T.~G.} \bibnamefont{Philbin}},
  \bibinfo{journal}{{New J. Phys.}} \textbf{\bibinfo{volume}{8}},
  \bibinfo{pages}{247} (\bibinfo{year}{2006}).

\bibitem[{\citenamefont{Pendry et~al.}(2006)\citenamefont{Pendry, Schurig, and
  Smith}}]{Pendry06}
\bibinfo{author}{\bibfnamefont{J.~B.} \bibnamefont{Pendry}},
  \bibinfo{author}{\bibfnamefont{D.}~\bibnamefont{Schurig}}, \bibnamefont{and}
  \bibinfo{author}{\bibfnamefont{D.~R.} \bibnamefont{Smith}},
  \bibinfo{journal}{{Science}} \textbf{\bibinfo{volume}{312}},
  \bibinfo{pages}{1780} (\bibinfo{year}{2006}).

\bibitem[{\citenamefont{Schurig et~al.}(2006)\citenamefont{Schurig, Mock,
  Justice, Cummer, Pendry, Starr, and Smith}}]{Schurig06}
\bibinfo{author}{\bibfnamefont{D.}~\bibnamefont{Schurig}},
  \bibinfo{author}{\bibfnamefont{J.~J.} \bibnamefont{Mock}},
  \bibinfo{author}{\bibfnamefont{B.~J.} \bibnamefont{Justice}},
  \bibinfo{author}{\bibfnamefont{S.~A.} \bibnamefont{Cummer}},
  \bibinfo{author}{\bibfnamefont{J.~B.} \bibnamefont{Pendry}},
  \bibinfo{author}{\bibfnamefont{A.~F.} \bibnamefont{Starr}}, \bibnamefont{and}
  \bibinfo{author}{\bibfnamefont{D.~R.} \bibnamefont{Smith}},
  \bibinfo{journal}{{Science}} \textbf{\bibinfo{volume}{314}},
  \bibinfo{pages}{977} (\bibinfo{year}{2006}).

\bibitem[{\citenamefont{{A. Greenleaf, M. Lassas, and G.
  Uhlmann}}(2003{\natexlab{a}})}]{Greenleaf2}
\bibinfo{author}{\bibnamefont{{A. Greenleaf, M. Lassas, and G. Uhlmann}}},
  \bibinfo{journal}{{Math. Res. Lett.}} \textbf{\bibinfo{volume}{10}},
  \bibinfo{pages}{685} (\bibinfo{year}{2003}{\natexlab{a}});
  \bibinfo{journal}{{Physiol. Meas.}} \textbf{\bibinfo{volume}{24}},
  \bibinfo{pages}{413} (\bibinfo{year}{2003}{\natexlab{b}}).


\bibitem[{\citenamefont{{J. Li and J. B. Pendry}}(2008)}]{PendryLi}
\bibinfo{author}{\bibnamefont{{J. Li and J. B. Pendry}}},
  \bibinfo{journal}{{Phys. Rev. Lett.}} \textbf{\bibinfo{volume}{101}},
  \bibinfo{pages}{203901} (\bibinfo{year}{2008}).

\bibitem[{\citenamefont{Pendry}(2009)}]{Pendry09}
\bibinfo{author}{\bibfnamefont{J.~B.} \bibnamefont{Pendry}},
  \bibinfo{journal}{{Nature}} \textbf{\bibinfo{volume}{460}},
  \bibinfo{pages}{579} (\bibinfo{year}{2009}).

\bibitem[{\citenamefont{{A. V. Kildishev, W. Cai, U. K. Chettiar, and V. M.
  Shalaev}}(2008)}]{Shalaev08}
\bibinfo{author}{\bibnamefont{{A. V. Kildishev, W. Cai, U. K. Chettiar, and V.
  M. Shalaev}}}, \bibinfo{journal}{{New J. Phys.}}
  \textbf{\bibinfo{volume}{10}}, \bibinfo{pages}{115029}
  (\bibinfo{year}{2008}).

\bibitem[{\citenamefont{Farhat et~al.}(2009)\citenamefont{Farhat, Guenneau, and
  Enoch}}]{Enoch09}
\bibinfo{author}{\bibfnamefont{M.}~\bibnamefont{Farhat}},
  \bibinfo{author}{\bibfnamefont{S.}~\bibnamefont{Guenneau}}, \bibnamefont{and}
  \bibinfo{author}{\bibfnamefont{S.}~\bibnamefont{Enoch}},
  \bibinfo{journal}{{Phys. Rev. Lett.}} \textbf{\bibinfo{volume}{103}},
  \bibinfo{pages}{024301} (\bibinfo{year}{2009}).

\bibitem[{\citenamefont{Valentine et~al.}(2009)\citenamefont{Valentine, Li,
  Zentgraf, Barta, and Zhang}}]{Zhang09}
\bibinfo{author}{\bibfnamefont{J.}~\bibnamefont{Valentine}},
  \bibinfo{author}{\bibfnamefont{J.}~\bibnamefont{Li}},
  \bibinfo{author}{\bibfnamefont{T.}~\bibnamefont{Zentgraf}},
  \bibinfo{author}{\bibfnamefont{G.}~\bibnamefont{Barta}}, \bibnamefont{and}
  \bibinfo{author}{\bibfnamefont{X.}~\bibnamefont{Zhang}},
  \bibinfo{journal}{{Nat. Mater.}} \textbf{\bibinfo{volume}{8}},
  \bibinfo{pages}{568} (\bibinfo{year}{2009}).

\bibitem[{\citenamefont{Lai et~al.}(2009{\natexlab{a}})\citenamefont{Lai, Chen,
  Zhang, and Chan}}]{Lai09}
\bibinfo{author}{\bibfnamefont{Y.}~\bibnamefont{Lai}},
  \bibinfo{author}{\bibfnamefont{H.}~\bibnamefont{Chen}},
  \bibinfo{author}{\bibfnamefont{Z.-Q.} \bibnamefont{Zhang}}, \bibnamefont{and}
  \bibinfo{author}{\bibfnamefont{C.~T.} \bibnamefont{Chan}},
  \bibinfo{journal}{{Phys. Rev. Lett.}} \textbf{\bibinfo{volume}{102}},
  \bibinfo{pages}{093901} (\bibinfo{year}{2009}{\natexlab{a}}).

\bibitem[{\citenamefont{Lai et~al.}(2009{\natexlab{b}})\citenamefont{Lai, Ng,
  Chen, Han, Xiao, Zhang, and Chan}}]{Lai09_2}
\bibinfo{author}{\bibfnamefont{Y.}~\bibnamefont{Lai}},
  \bibinfo{author}{\bibfnamefont{J.}~\bibnamefont{Ng}},
  \bibinfo{author}{\bibfnamefont{H.}~\bibnamefont{Chen}},
  \bibinfo{author}{\bibfnamefont{D.~Z.} \bibnamefont{Han}},
  \bibinfo{author}{\bibfnamefont{J.~J.} \bibnamefont{Xiao}},
  \bibinfo{author}{\bibfnamefont{Z.-Q.} \bibnamefont{Zhang}}, \bibnamefont{and}
  \bibinfo{author}{\bibfnamefont{C.~T.} \bibnamefont{Chan}},
  \bibinfo{journal}{{Phys. Rev. Lett.}} \textbf{\bibinfo{volume}{102}},
  \bibinfo{pages}{253902} (\bibinfo{year}{2009}{\natexlab{b}}).

\bibitem[{\citenamefont{{D. A. B. Miller}}(2006)}]{Miller06}
\bibinfo{author}{\bibnamefont{{D. A. B. Miller}}}, \bibinfo{journal}{{Opt.
  Express}} \textbf{\bibinfo{volume}{14}}, \bibinfo{pages}{12457}
  (\bibinfo{year}{2006}).

\bibitem[{\citenamefont{Vasquez
  et~al.}(2009{\natexlab{a}})\citenamefont{Vasquez, Milton, and
  Onofrei}}]{MiltonPRL}
\bibinfo{author}{\bibfnamefont{F.~G.} \bibnamefont{Vasquez}},
  \bibinfo{author}{\bibfnamefont{G.~W.} \bibnamefont{Milton}},
  \bibnamefont{and} \bibinfo{author}{\bibfnamefont{D.}~\bibnamefont{Onofrei}},
  \bibinfo{journal}{Phys. Rev. Lett.} \textbf{\bibinfo{volume}{103}},
  \bibinfo{pages}{073901} (\bibinfo{year}{2009}{\natexlab{a}}).

\bibitem[{\citenamefont{{G. W. Milton and N. A. P.
  Nicorovici}}(2006)}]{Milton06}
\bibinfo{author}{\bibnamefont{{G. W. Milton and N. A. P. Nicorovici}}},
  \bibinfo{journal}{{Proc. R. Soc. A}} \textbf{\bibinfo{volume}{462}},
  \bibinfo{pages}{3027} (\bibinfo{year}{2006}).

\bibitem[{\citenamefont{{N. P. Nicorovici, G. W. Milton, R. C. McPhedran, and
  L. C. Botten}}(2007)}]{Nicorovici07}
\bibinfo{author}{\bibnamefont{{N. A. Nicorovici, G. W. Milton, R. C. McPhedran,
  and L. C. Botten}}}, \bibinfo{journal}{{Opt. Express}}
  \textbf{\bibinfo{volume}{15}}, \bibinfo{pages}{6314} (\bibinfo{year}{2007}).

\bibitem[{\citenamefont{Vasquez
  et~al.}(2009{\natexlab{b}})\citenamefont{Vasquez, Milton, and
  Onofrei}}]{MiltonOE}
\bibinfo{author}{\bibfnamefont{F.~G.} \bibnamefont{Vasquez}},
  \bibinfo{author}{\bibfnamefont{G.~W.} \bibnamefont{Milton}},
  \bibnamefont{and} \bibinfo{author}{\bibfnamefont{D.}~\bibnamefont{Onofrei}},
  \bibinfo{journal}{Opt. Express} \textbf{\bibinfo{volume}{17}},
  \bibinfo{pages}{14800} (\bibinfo{year}{2009}{\natexlab{b}}).

\bibitem[{\citenamefont{{J. J. Xiao and C. T. Chan}}(2008)}]{Xiao08}
\bibinfo{author}{\bibnamefont{{See for example, J. J. Xiao and C. T. Chan}}},
  \bibinfo{journal}{{J. Opt. Soc. Am. B}} \textbf{\bibinfo{volume}{25}},
  \bibinfo{pages}{1553} (\bibinfo{year}{2008}), and refereces there in.

\bibitem[{\citenamefont{Kong}(2005)}]{JAKong}
\bibinfo{author}{\bibfnamefont{J.~A.} \bibnamefont{Kong}},
  \emph{\bibinfo{title}{{Electromagnetic Wave Theory}}}
  (\bibinfo{publisher}{EMW Publishing}, \bibinfo{address}{Cambridge,
  Massachusetts, USA}, \bibinfo{year}{2005}).

\end{thebibliography}

\end{document}